\documentclass[twocolumn,pra,amsmath,amssymb]{revtex4}
\usepackage{graphicx}
\usepackage{amssymb,amsthm,amsfonts,amstext}
\usepackage{url}
\def\be{\begin{equation}}
\def\ee{\end{equation}}
\def\bea{\begin{eqnarray}}
\def\eea{\end{eqnarray}}
\def\bma{\begin{mathletters}}
\def\ema{\end{mathletters}}

\def\0{\overline{0}}

\def\q0{\underline{0}}

\def\B{{\cal B}}

\def\one{\leavevmode\hbox{\small1\normalsize\kern-.33em1}}

\begin{document}

\title{Multisetting Bell-type inequalities for detecting genuine tripartite entanglement}

\author{K\'aroly F. P\'al}
\email{kfpal@atomki.hu}
\affiliation{Institute of Nuclear Research of the Hungarian Academy of Sciences\\
H-4001 Debrecen, P.O. Box 51, Hungary}

\author{Tam\'as V\'ertesi}
\email{tvertesi@dtp.atomki.hu}
\affiliation{Institute of Nuclear Research of the Hungarian Academy of Sciences\\
H-4001 Debrecen, P.O. Box 51, Hungary}

\date{\today}

\begin{abstract}
In a recent paper, Bancal et al.~put forward the concept of
device-independent witnesses of genuine multipartite entanglement.
These witnesses are capable of verifying genuine multipartite
entanglement produced in a lab without resorting to any knowledge of the
dimension of the state space or of the specific form of the measurement operators. As a
by-product they found a three-party three-setting Bell inequality
which enables to detect genuine tripartite entanglement in a noisy 3-qubit
Greenberger-Horne-Zeilinger (GHZ) state for visibilities as low as
$2/3$ in a device-independent way. In this paper, we generalize this inequality 
to an arbitrary number of settings, demonstrating a threshold visibility of
$2/\pi\sim0.6366$ for number of settings going to infinity. We also
present a pseudo-telepathy Bell inequality achieving the
same threshold value. We argue that our device-independent witnesses
are optimal in the sense that the above value cannot be beaten with
three-party-correlation Bell inequalities.
\end{abstract}

\maketitle

\section{Introduction}
Quantum theory allows correlations between remote systems, which
are fundamentally different from classical correlations
\cite{Bell}. Quantum entanglement is in the heart of this
phenomenon \cite{HHHH}. Already two entangled particles give rise
to correlations not reproducible within any local realistic theory
\cite{CHSH}. However, moving to more particles a much richer
structure and various types of entanglement arise \cite{Acin}
suggesting novel applications such as quantum computation using
cluster states \cite{qcluster}, sub-shotnoise metrology
\cite{qmetrology}, or multiparty quantum networking \cite{qcomm}.
In these tasks, genuinely entangled particles offer enhanced performance.
Hence, it is a central problem to decide whether in an actual experiment
genuinely multipartite entanglement has been produced, or
alternatively, the entangled state prepared in the laboratory could be explained
without requiring the interaction of all particles. In the latter
case, we say that the state created is biseparable. Focussing on
the tripartite case, a biseparable state $\rho_{bs}$ can be
written as
\begin{equation}
\rho_{bs}=\sum_i{p_i|\phi_i\rangle\langle \phi_i|},
\label{eq:bs}
\end{equation}
where the pure states $\phi_i$ are separable with respect to one
of the three bipartitions $1|23$, $12|3$, $13|2$, and the weights
$p_i>0$ add up to 1. For more than three parties, the generalization is straightforward.

Several experiments have been conducted so far generating
multipartite entangled photonic states up to six photons (for
instance, Ref.~\cite{six} generated a Dicke state of six
photons). One of the traditional approaches to decide on the
existence of genuine multipartite entanglement consists in
performing a complete state tomography, and then deducing the kind
of entanglement directly from the density matrix using witness
operators. Alternatively, the experimentalist may measure
cleverly chosen witness operators, thereby reducing the number of
correlation terms to be measured in the actual experiment
\cite{four}. However, a common drawback is that in both cases
the experimentalist needs to have a precise control over the system
on which the measurements are performed.

Remarkably, there is another route, avoiding the above problem,
building on the seminal work of John Bell \cite{Bell}: Bell
expressions are linear functions of joint correlations enabling one to
say important things in a black box scenario about the dimension of the systems,
the states involved, or the kind of measurements performed.
In particular, it is possible to decide on the presence of genuine
multipartite entanglement based on merely statistical data (that is, without relying
on any knowledge of the implementation of the devices involved in the measurement process) \cite{diew,BGLP}:
if a Bell value, coming from the statistics of a Bell experiment, is bigger
than a certain value achievable with measurements acting on
biseparable quantum states, then we can be sure that the state in
question is genuinely multipartite entangled. This approach has
been formalized more recently by Bancal et al.~\cite{BGLP},
coining the term device-independent witnesses of genuine multipartite entanglement
witnesses for such Bell expressions (for more details we refer the reader to that paper).

As a simplest illustration of a device-independent witness of genuine tripartite
entanglement, let us represent the Mermin polynomial~\cite{Mermin} in terms of three-party correlators,
\begin{align}
I_2\equiv& \langle \hat A_0\otimes\hat B_0\otimes\hat C_0\rangle-\langle
\hat A_0\otimes\hat B_1\otimes\hat C_1\rangle-\langle \hat A_1\otimes\hat B_0\otimes\hat C_1\rangle\nonumber\\&-\langle\hat A_1\otimes\hat B_1\otimes\hat C_0\rangle,
\label{eq:Mermin}
\end{align}
where $\langle \hat A_{\alpha}\otimes\hat B_{\beta}\otimes\hat C_{\gamma} \rangle$ designate the expected value of the product of three $\pm 1$-observables, $\hat A_{\alpha}$, $\hat B_{\beta}$, $\hat C_{\gamma}$.
It has been shown in Ref.~\cite{diew}, that $I_2\le 2\sqrt 2$ for
biseparable quantum states (${\cal B}_2=2\sqrt 2$), whereas the maximum quantum value saturates the
algebraic limit of 4 (${\cal Q}_2=4$), hence the violation of the
bound ${\cal B}_2$ implies genuine tripartite entanglement. Note that this reasoning holds true independently on the size of the Hilbert space dimension or on the type of measurements carried out. Hence, Mermin inequality serves as a device-independent witness of genuine tripartite entanglement~\cite{BGLP}.
Let us now take the noisy 3-qubit GHZ state,
\begin{equation}
\rho(V)=V|GHZ\rangle\langle GHZ|+(1-V)\frac{\one}{8},
\label{eq:noisyGHZ}
\end{equation}
where
$|GHZ\rangle=\left(|000\rangle+|111\rangle\right)/\sqrt 2$ is the 3-qubit GHZ state~\cite{GHZ}, and
$V$ is the visibility parameter. The measurements achieving the
bounds ${\cal Q}_2$ and ${\cal B}_2$ correspond to traceless observables,
entailing the threshold visibility $V={\cal B}_2/{\cal Q}_2=1/\sqrt 2$. Hence, genuine tripartite entanglement in the noisy GHZ state (for $V>1/\sqrt 2$) can be detected in a device-independent way.

More recently, however, Bancal et al.~\cite{BGLP} managed to lower the threshold visibility of the noisy 3-party GHZ state to $V=2/3$ by considering a three-party three-setting Bell inequality, which can be considered as a three-setting generalization of the two-setting Mermin inequality. Note that similarly to the Mermin inequality, the Bancal et al.~inequality extends to more than three parties as well \cite{BGLP}.

In the present paper, we generalize the three-setting three-party Bancal et al.~inequality to an arbitrary
number of settings $m$, exhibiting the threshold visibility $V=1/(m\sin(\pi/2m))$, which approaches $V=2/\pi$ for large number of settings. This generalization is discussed in section~\ref{ketto}, whereas another family of Bell inequalities, based on the extended parity game~\cite{parity}, is discussed in section~\ref{harom}. Notably, this game exhibits pseudo-telepathy \cite{PT}, and the corresponding Bell inequality has the same performance (for $m$ a power of 2) as our Bell inequality of section~\ref{ketto}.

\section{Multisetting tripartite Bell-type~inequalities}\label{ketto}

Let us introduce the $m$-setting tripartite Bell expression,
\begin{equation}
I_m=\sum_{\alpha,\beta,\gamma=0}^{m-1}{M_{\alpha\beta\gamma}\langle \hat A_{\alpha}\otimes\hat B_{\beta}\otimes\hat C_{\gamma}\rangle},
\label{Im}
\end{equation}
where the matrix of Bell coefficients is defined by
\begin{equation}
M_{\alpha\beta\gamma}=\cos\left[\frac{\pi}{m}(\alpha+\beta+\gamma-\Delta)\right],
\label{eq:Bellcont}
\end{equation}
where indices $\alpha$, $\beta$ and $\gamma$ may take the values
of $0,1,\dots,m-1$, and $\Delta$ may be any real number. By choosing $m=2$ and $\Delta=0$, the Mermin polynomial~(\ref{eq:Mermin}) is recovered. On the other hand, for $m=3$ and $\Delta=-1/2$, we obtain the polynomial of Bancal et al.~\cite{BGLP} (apart from an irrelevant multiplicative factor).

We next exhibit a lower bound on the quantum maximum, ${\cal Q}^l_m=m^3/2$, as a
function of number of settings $m$. Then an upper bound is given on the biseparable quantum maximum, which is shown to be attained by von Neumann-type projective measurements, ${\cal B}_m=m^2/(2\sin(\pi/2m))$. This implies the threshold visibility $V={\cal B}_m/{\cal Q}_m\le {\cal B}_m/{\cal Q}^l_m = 1/(m\sin(\pi/2m))$ tending to $2/\pi$ in the limit of large number of measurement settings.

We wish to note that Bancal et al.~(Appendix~C in~\cite{BGLP}) presented a biseparable model simulating all the single-party expectations $\langle \hat A\rangle$, $\langle \hat B\rangle$, $\langle \hat C\rangle$, two-party correlators $\langle \hat A\otimes\hat B\rangle$, $\langle \hat A\otimes\hat C\rangle$, $\langle \hat B\otimes\hat C\rangle$ and three-party correlators $\langle \hat A\otimes\hat B\otimes\hat C\rangle$, achievable with von Neumann measurements on the noisy 3-qubit GHZ state~(\ref{eq:noisyGHZ}) of visibility $V\le 1/2$. Within this biseparable model all of the three parties may share local random variables, but at most two parties can share a quantum state at a given time. It can be shown that if we are content with simulating only the three-party correlators, then the threshold visibility becomes a higher value, $V=2/\pi$. This implies that it is not possible to detect genuine tripartite entanglement in the 3-qubit GHZ state in the range $V\le 2/\pi$ by applying von Neumann measurements and considering Bell expressions which are sums of three-party correlators. In this sense, our family of Bell inequalities is optimal, giving $V\rightarrow2/\pi$ when $m$ goes to infinity.

\vspace{10pt}
\noindent\textbf{Lower bound on the quantum maximum, ${\cal Q}^l_m $.}
If each of the participants performs a von Neumann projective measurement on one component of
a shared 3-qubit GHZ state, the tripartite correlation of their
measurement values can be written as (see for instance Appendix~C in Ref.~\cite{BGLP}):
\begin{equation}
\langle\hat A_\alpha\otimes\hat B_\beta\otimes\hat
C_\gamma\rangle=\sin\theta^A_\alpha\sin\theta^B_\beta
\sin\theta^C_\gamma\cos(\varphi^A_\alpha+\varphi^B_\beta+\varphi^C_\gamma),
\label{eq:3corr}
\end{equation}
where the $\alpha$th, $\beta$th and $\gamma$th measurement
operator $\hat A_\alpha$, $\hat B_\beta$ and $\hat C_\gamma$ of
Alice, Bob and Cecil, respectively are given as:
\begin{align}
\hat
A_\alpha&=\cos\varphi^A_\alpha\sin\theta^A_\alpha\hat\sigma_x+
\sin\varphi^A_\alpha\sin\theta^A_\alpha\hat\sigma_y+\cos\theta^A_\alpha\hat\sigma_z\nonumber\\
\hat B_\beta&=\cos\varphi^B_\beta\sin\theta^B_\beta\hat\sigma_x+
\sin\varphi^B_\beta\sin\theta^B_\beta\hat\sigma_y+\cos\theta^B_\beta\hat\sigma_z\nonumber\\
\hat
C_\gamma&=\cos\varphi^C_\gamma\sin\theta^C_\gamma\hat\sigma_x+
\sin\varphi^C_\gamma\sin\theta^C_\gamma\hat\sigma_y+\cos\theta^C_\gamma\hat\sigma_z,
\label{eq:measoper}
\end{align}
where $\hat\sigma_x$, $\hat\sigma_y$ and $\hat\sigma_z$ are the
Pauli operators.

With the choice of $\theta^A_\mu=\theta^B_\mu=\theta^C_\mu=0$ and
$\varphi^A_\mu=\varphi^B_\mu=\varphi^C_\mu= \pi(\mu-\Delta/3)/m$
each tripartite correlation will take the same value as the Bell
coefficient to be multiplied with, and the quantum value of the
Bell expression will be easy to calculate:
\begin{align}
{\cal
Q}^l_m&=\sum_{\alpha=0}^{m-1}\sum_{\beta=0}^{m-1}\sum_{\gamma=0}^{m-1}M_{\alpha\beta\gamma}
\langle\hat A_\alpha\otimes\hat B_\beta\otimes\hat C_\gamma\rangle\nonumber\\
&=\sum_{\alpha=0}^{m-1}\sum_{\beta=0}^{m-1}\sum_{\gamma=0}^{m-1}
\cos^2\left[\frac{\pi}{m}(\alpha+\beta+\gamma-\Delta)\right]\nonumber\\
&=\frac{1}{2}\sum_{\alpha=0}^{m-1}\sum_{\beta=0}^{m-1}\sum_{\gamma=0}^{m-1}\left\{
1-\cos\left[\frac{2\pi}{m}(\alpha+\beta+\gamma-\Delta)\right]\right\}\nonumber\\
&=\frac{m^3}{2}. \label{eq:qbound}
\end{align}

This value ${\cal Q}^l_m$ is a lower bound for the maximum quantum
value ${\cal Q}_m$.

For the maximum of the biseparable value first we will give an upper bound
(${\cal B}^u_m$), then we will prove that this bound can be saturated, that is,
(${\cal B}^l_m={\cal B}^u_m={\cal B}_m$).

\vspace{10pt}
\noindent\textbf{Upper bound on the biseparable quantum value, ${\cal B}^u_m $.}
The value to be calculated is,
\begin{align}
{{\cal B}_m}&=\max\sum_{\alpha=0}^{m-1}\sum_{\beta=0}^{m-1}\sum_{\gamma=0}^{m-1}M_{\alpha\beta\gamma}
A_\alpha\langle\hat B_\beta\otimes\hat C_\gamma\rangle\nonumber\\
&=\max\sum_{\beta=0}^{m-1}\sum_{\gamma=0}^{m-1}M^{\{A\}}_{\beta\gamma}\langle\hat
B_\beta\otimes\hat C_\gamma\rangle, \label{eq:bisep}
\end{align}
where we used the fact that Bell inequality~(\ref{Im})
is linear in the correlators, and that a biseparable density
matrix~(\ref{eq:bs}) is a convex combination of pure states, hence
it is enough to take the 3-party correlators in the form $\langle
\hat A_{\alpha}\rangle\langle \hat B_{\beta}\otimes \hat C_{\gamma}\rangle$. In
Eq.~(\ref{eq:bisep}) each of $A_\alpha$ may take the value of either
$+1$ or $-1$, Bob and Cecil may share any quantum state and
perform measurements on them, the operators of their measurement
settings are $\hat B_\beta$ and $\hat C_\gamma$, respectively, and
the coefficients of the two-partite Bell inequality, which depends
on the actual choice of $A_\alpha$ are:
\begin{equation}
M^{\{A\}}_{\beta\gamma}=\sum_{\alpha=0}^{m-1}A_\alpha\cos\left[\frac{\pi}{m}(\alpha+\beta+\gamma-\Delta)\right].
\label{eq:Bellbpart}
\end{equation}
We note that due to the symmetry of the Bell expression under
party exchange, it is enough to consider the case when it is Alice
who may not share an entangled quantum object with the others.

From the work of Ref.~\cite{Wehner} it easily follows that for bipartite correlation type
Bell inequalities with an equal number of measurement settings per
party, an upper bound for the maximum quantum value is the largest
of the singular values of the matrix defined by the Bell
coefficients multiplied by the number of measurement settings. In
the present case the matrix depends on the sum of its indices.
Therefore,
$M^{\{A\}}_{(\beta+1)\gamma}=M^{\{A\}}_{\beta(\gamma+1)}$, that is
each row contains the elements of the preceding row, shifted to
the left. From Eq.~(\ref{eq:Bellbpart}) it is also clear, that
$M^{\{A\}}_{(\beta+1)m}=-M^{\{A\}}_{\beta 1}$, that is the last
element of each row is the same as minus one times the first
element of the preceding row. These properties are very similar to
the properties defining circulant matrices \cite{circulant}, whose
eigenvectors are independent of the actual values of its elements,
and therefore whose eigenvalues are very easy to derive. There
are just two differences. In the case of the circulant matrices
the elements are shifted not to the left, but to the right.
Furthermore, they do it cyclically, that is there is no change of
sign when the last element takes the first place in the next row.
The first difference is easily corrected if we rearrange Cecil's
measurement settings into the opposite order. Fortunately, the
change of sign of the matrix element poses no serious problem
either, because it can be shown that the eigenvectors of these
modified circulant matrices are also independent of the actual
values of the elements of the matrix, they are given as:
\begin{align}
v_j&=\left(1,\omega_j,\omega_j^2,\dots\omega_j^{m-1}\right)^T,\nonumber\\
\omega_j&=e^{\frac{2\pi i(j+1/2)}{m}}, \label{eq:eigvecmcir}
\end{align}
where $j=0,\dots,m-1$. The difference from the circulant case
\cite{circulant} is the $1/2$ term in the exponent. To calculate
the eigenvalues we only need the first row of the matrix.
Therefore, we get the upper bound for the biseparable value as:
\begin{equation}
{\cal B}^u_m=\max
m\left|\sum_{\alpha=0}^{m-1}\sum_{\gamma=0}^{m-1}A_\alpha
\cos\left[\frac{\pi(\alpha-\Delta'-\gamma)}{m}\right]\omega_j^\gamma\right|,
\label{eq:bisepu1}
\end{equation}
where we introduced the notation $\Delta'\equiv\Delta-m+1$. By
substituting the $\omega_j$ from Eq.~(\ref{eq:eigvecmcir}) and
using identity
\begin{align}
\cos\frac{\pi(\alpha-\Delta'-\gamma)}{m}=&\cos\frac{\pi(\alpha-\Delta')}{m}\cos\frac{\pi\gamma}{m}\nonumber\\
&+\sin\frac{\pi(\alpha-\Delta')}{m}\sin\frac{\pi\gamma}{m}
\label{eq:idenc}
\end{align}
we arrive at:
\begin{align}
{\cal B}^u_m=\max m\Bigg|\sum_{\alpha=0}^{m-1}A_\alpha
\Big[&\cos\frac{\pi(\alpha-\Delta')}{m}(D_1^j+iD_2^j)\nonumber\\
&+\sin\frac{\pi(\alpha-\Delta')}{m}(D_3^j+iD_4^j)\Big]\Bigg|,
\label{eq:bisepu2}
\end{align}
where
\begin{align}
D_1^j&=\sum_{\gamma=0}^{m-1}\cos\frac{\pi\gamma}{m}\cos\frac{2\pi(j+1/2)\gamma}{m}\nonumber\\
D_2^j&=\sum_{\gamma=0}^{m-1}\cos\frac{\pi\gamma}{m}\sin\frac{2\pi(j+1/2)\gamma}{m}\nonumber\\
D_3^j&=\sum_{\gamma=0}^{m-1}\sin\frac{\pi\gamma}{m}\cos\frac{2\pi(j+1/2)\gamma}{m}\nonumber\\
D_4^j&=\sum_{\gamma=0}^{m-1}\sin\frac{\pi\gamma}{m}\sin\frac{2\pi(j+1/2)\gamma}{m}.
\label{eq:D1234}
\end{align}
However,
\begin{equation}
D_2^j\pm D_3^j=\sum_{\gamma=0}^{m-1}\sin\frac{2\pi(j+1/2\pm
1/2)\gamma}{m}=0,
\end{equation}
therefore $D_2^j=D_3^j=0$, and
\begin{equation}
D_1^j\pm D_4^j=\sum_{\gamma=0}^{m-1}\cos\frac{2\pi(j+1/2\mp
1/2)\gamma}{m},
\end{equation}
from which it follows that $D_1^j=D_4^j=0$ for $1\leq j\leq m-2$,
$D_1^0=D_4^0=m/2$, and $D_1^{m-1}=-D_4^{m-1}=m/2$. To get the
maximum we must take either $j=0$ or $j=m-1$. For $j=0$ we get:
\begin{equation}
{\cal B}^u_m=\max\frac{m^2}{2}\left|\sum_{\alpha=0}^{m-1}A_\alpha
e^{\frac{i\pi(\alpha-\Delta')}{m}}\right|. \label{eq:bisepu3}
\end{equation}
If we have taken $j=m-1$ instead of $j=0$, we would have got the
complex conjugate of the numbers whose absolute value has to be
taken, which would have given the same result. In
Eq.~(\ref{eq:bisepu3}) we have to add $m$ vectors on the complex
plane, each pointing towards corners of a regular polygon of $2m$
sides, and then we have to take the length of this vector. Each
vector lies on a different diagonal of the polygon, but may point
towards either direction depending on the value of $A_\alpha$. It
can be shown that we get the largest value if the vectors taken in
some order point towards consecutive corners. All such
arrangements give obviously the same result. We get one of those
arrangements if we take $A_\alpha=1$. The result does not depend
on $\Delta'$, as changing $\Delta'$ means only an overall rotation
of the arrangement. Let us take $\Delta'=-1/2$. Then the set of
numbers will be symmetric with respect to the imaginary axis,
therefore the real part of the sum will be zero, while the
imaginary part will be positive. Then we get
\begin{align}
{\cal B}^u_m=&\frac{m^2}{2}\sum_{\alpha=0}^{m-1}\sin{\frac{\pi(\alpha+1/2)}{m}}\nonumber\\
=&\frac{m^2}{2\sin\frac{\pi}{2m}}\sum_{\alpha=0}^{m-1}
\Big(\sin\frac{\pi(\alpha+1/2)}{m}\sin\frac{\pi}{2m}\nonumber\\
&+\cos\frac{\pi(\alpha+1/2)}{m}\cos\frac{\pi}{2m}\Big)\nonumber\\
=&\frac{m^2}{2\sin\frac{\pi}{2m}}\sum_{\alpha=0}^{m-1}\cos\frac{\pi\alpha}{m}\nonumber\\
=&\frac{m^2}{2\sin\frac{\pi}{2m}}. \label{eq:bisepu4}
\end{align}
Here we have used that
$\sum_{\alpha=0}^{m-1}\cos[\pi(\alpha+1/2)/m]=0$, and that
$\cos(\pi\alpha/m)= -\cos[\pi(m-\alpha)/m]$.

Now we will show that this upper bound can be saturated.

\vspace{10pt}
\noindent\textbf{Lower bound on the biseparable quantum value, ${\cal B}^l_m$.}
If $\sum_{\beta=0}^{m-1}\sum_{\gamma=0}^{m-1} \bar
M_{\beta\gamma}\vec C_\beta\cdot\vec C_\gamma$ is a certain
number, where $\vec C_\beta$ and $\vec C_\gamma$ are Euclidean
unit vectors, then there exist measurement operators giving
the same number as the quantum value of the bipartite correlation
type Bell inequality of coefficients $\bar M_{\beta\gamma}$,
applied on the maximally entangled state \cite{Tsirelson}. In case
of two dimensional vectors pairs of real qubits are sufficient.
Let $\bar M_{\beta\gamma}\equiv M^{\{A\}}_{\beta\gamma}$ with all
$A_\alpha=+1$ (see Eq.~(\ref{eq:Bellbpart})), let Cecil's vectors be
\begin{equation}
\vec C_\gamma=\left(\begin{array}{c}\cos\frac{\pi\gamma}{m}\\
\sin\frac{\pi\gamma}{m}\end{array}\right), \label{eq:Vecc}
\end{equation}
and let us choose $\vec B_\beta$ optimally, that is $\vec
B_\beta=\sum_{\gamma=0}^{m-1}\bar M_{\beta\gamma}\vec
C_\gamma/|\sum_{\gamma=0}^{m-1}\bar M_{\beta\gamma}\vec
C_\gamma|$. Then the corresponding quantum value is:
\begin{align}
{\cal B}^l_m&=\sum_{\beta=0}^{m-1}\left|\sum_{\gamma=0}^{m-1}\bar M_{\beta\gamma}\vec C_\gamma\right|\nonumber\\
&=\sum_{\beta=0}^{m-1}\left|\sum_{\alpha=0}^{m-1}\sum_{\gamma=0}^{m-1}\cos\frac{\pi(\alpha+\beta+\gamma-\Delta)}{m}
\left(\begin{array}{c}\cos\frac{\pi\gamma}{m}\\
\sin\frac{\pi\gamma}{m}\end{array}\right)\right|.
\label{eq:bisepl}
\end{align}
Now if we follow analogous steps to the ones we have taken
calculating the value of ${\cal B}^u$, we will arrive at the same
result. We can also easily see this if we compare
Eq.~(\ref{eq:bisepl}) to Eq.~(\ref{eq:bisepu1}). The maximum value
of the latter expression has been attained with $A_\alpha=1$ and
$j=0$. By substituting these values, and also $\omega_0$ from
Eq.~(\ref{eq:eigvecmcir}), we get almost the same formula as
Eq.~(\ref{eq:bisepl}), indeed. The $\omega_0^\gamma$ complex
numbers correspond to the same vectors on the complex plane as the
two dimensional vectors appearing in Eq.~(\ref{eq:bisepl}). From
the calculation of ${\cal B}^u$ it turns out that the value does
not depend on $\Delta'$, so the absolute value in
Eq.~(\ref{eq:bisepl}) does not depend on $\Delta-\beta$ either,
therefore, we may replace the summation in terms of $\beta$ for a
multiplicative factor of $m$. The only remaining difference is the
opposite sign of $\gamma$ in the cosine, but that will not affect
the result either.

As the upper and lower bound for the biseparable case are equal,
the biseparable value itself is given by Eq.~(\ref{eq:bisepu4}).
For the quantum value we have only proven a lower bound (see
Eq.~(\ref{eq:qbound})). Therefore, for this family of Bell
inequalities the ratio of the quantum and the biseparable values
(which equals the visibility threshold) satisfies:
\begin{equation}
V=\frac{{\cal B}_m}{{\cal Q}_m}\leq \frac{1}{m\sin\frac{\pi}{2m}}. \label{eq:QperB}
\end{equation}
We believe that the lower bound ${\cal Q}^l_m$ we have given in (\ref{eq:qbound}) is actually the quantum maximum itself, and the above expression is valid as an equality.

\section{Bell-type inequalities based on the extended parity game}\label{harom}

Now we define another family of multisetting tripartite inequalities
giving the same ratio of ${{\cal B}_m}/{{\cal Q}_m}$ as the right hand
side of Eq.~(\ref{eq:QperB}), at least when $m$ is a power of 2.

The Bell coefficients may only take the values of zero, one and
minus one, namely:
\begin{equation}
M_{\alpha\beta\gamma}=\left\{\begin{array}{ll}
0 &\text{if } (\alpha+\beta+\gamma) \mod m\neq 0,\\
1 &\text{if } (\alpha+\beta+\gamma)/m \text{ is even},\\
-1 &\text{if } (\alpha+\beta+\gamma)/m \text{ is odd},\\
\end{array}
\right.
\label{eq:BellPT}
\end{equation}
and $\alpha,\beta,\gamma=1,\dots,m-1$. These Bell coefficients correspond to the so-called
extended parity game considered in Ref.~\cite{parity}.
An equivalent definition, more similar to the definition of the Bell inequality~(\ref{eq:Bellcont}) treated in section~\ref{ketto} is that
$M_{\alpha\beta\gamma}=\cos[\pi(\alpha+\beta+\gamma)/m]$, whenever the absolute value of this expression is one, and $M_{\alpha\beta\gamma}=0$ otherwise.

\vspace{10pt}
\noindent\textbf{Maximum quantum value, ${\cal\breve Q}_m$.}
We get a lower bound for the quantum value with measurement operators given in
Eq.~(\ref{eq:measoper}) applied to components of a 3-qubit GHZ
state, with $\theta^A_\mu=\theta^B_\mu=\theta^C_\mu=0$ and
$\varphi^A_\mu=\varphi^B_\mu=\varphi^C_\mu= \pi\mu/m$. Using
Eq.~(\ref{eq:3corr}) it is clear that each nonzero Bell
coefficient will be multiplied by the same value as itself,
therefore, the quantum value will be equal to the sum of the
absolute values of the Bell coefficients, that is with the no
signalling limit, which is an upper bound for the quantum value. Hence it has the property of pseudo-telepathy \cite{PT}.
From this it follows, that the quantum value will be nothing else
than the number of nonzero Bell coefficients, which is actually
$m^2$. To see this, it is enough to note that however we slice up
the $m\times m\times m$ arrangement, each resulting $m\times m$
matrices will have exactly one nonzero number (plus or minus one)
in each of its rows and columns. We can get such a row or column
by fixing two of the indices of $M_{\alpha\beta\gamma}$. The sum
of the indices we get this way are $m$ consecutive numbers,
exactly one of them will be divisible by $m$. Such a matrix will
have $m$ nonzero elements, the $m$ slices together will contain
$m^2$ such elements, therefore the quantum value and the no
signalling limit will be ${\cal\breve Q}_m=m^2$.

We now place an upper bound on the maximum of the biseparable value (${\cal\breve\B}^l_m$), and then we prove that this bound can be saturated, that is, (${\cal\breve B}^l_m={\cal\breve B}_m$).

\vspace{10pt}
\noindent\textbf{Upper bound on the biseparable quantum value, ${\cal\breve B}^u_m$.}
A further property of the slices of the present $m\times m\times
m$ arrangement is that they are modified circulant matrices like
in the case of the previous family, which can be shown exactly the
same way as we have shown there. To get the matrices relevant to
the biseparable value, we have to add up the slices with different
signs. If it is Alice who is not allowed to share entangled state
with the others, this sum is
$M^{\{A\}}_{\beta\gamma}=\sum_{\alpha=0}^{m-1}A_\alpha
M_{\alpha\beta\gamma}$. Due to the property of the arrangement,
for each matrix element, all terms of the sum but one will be
zero. Therefore, each entry of $M^{\{A\}}_{\beta\gamma}$ will
either be one or minus one. Moreover, this matrix will also be a
modified circulant one, and its first line, which determines all
the others, may contain any combination of plus and minus one
values, depending on $A_\alpha$. Let
$a_{m-1-\gamma}\equiv\sum_{\alpha=0}^{m-1}A_\alpha M_{\alpha
0\gamma}$, that is the first line of $M^{\{A\}}_{0\gamma}$ written
in opposite order. Then an upper bound for the biseparable value
may be written as:
\begin{equation}
{\cal\breve B}^u_m=\max m\left|\sum_{\gamma=0}^{m-1}a_\gamma
e^{\frac{i\pi(2j+1)\gamma}{m}}\right|, \label{eq:bisepuns}
\end{equation}
Let us consider the case of $j=0$. Then what we get is the same as
Eq.~(\ref{eq:bisepu3}) but with $\Delta'=0$ (which is irrelevant),
and a prefactor of $m$ instead of $m^2/2$. Then, according to
Eq.~(\ref{eq:bisepu4}), the result is $m/\sin(\pi/2m)$. We will
show that this actually is the upper bound, whenever $m$ is a
power of two. As we have discussed earlier, $\exp(i\pi\gamma/m)$,
which corresponds to $j=0$, will point towards consecutive corners
of a regular polygon of $2m$ edges on the complex plane while
$\gamma$ takes all values between zero and $m-1$. If $m$ is a
power of two, then for any $j$, $\exp(i\pi (2j+1)\gamma/m)$ will
point towards different corners for the different $\gamma$ values,
moreover if one of them will point towards one corner, there will
be none pointing towards the opposite corner. The reason is that
$(2j+1)\gamma$ is never divisible with $m$ in this case. Choosing
$a_\gamma$ appropriately one can achieve that the terms to be
added point towards consecutive corners, if taken in some order, which maximizes the
absolute value of the sum. This is not true if $m$ is divisible
with an odd number. When $2j+1$ is equal to this number, for
$\gamma=m/(2j+1)$ the value of $\exp(\pi i(2j+1)\gamma/m)=-1$,
which lies opposite to $+1$, the value for $\gamma=0$. In this
case not all corners can be reached with appropriate choices of
$a_\gamma$, and the other corners can be reached more than once, and
${\cal\breve B}^u_m$ may be larger than what we have
calculated. If $m$ is odd, for $2j+1=m$ with $a_\gamma=-1^\gamma$
we even reach the no signalling limit.

Now we show that we can actually reach the value of ${\cal\breve B}^u_m=m/\sin(\pi/2m)$.

\vspace{10pt}
\noindent\textbf{Lower bound on the biseparable quantum value, ${\cal\breve B}^l_m$.}
The coefficients of the reduced Bell inequality
are the elements of the modified circulant matrix $\tilde
M_{\beta\gamma}$ whose entries in the first line are all $+1$. The
appropriate Euclidean vectors $\vec C_\gamma$ are the same as the
ones already defined in Eq.~(\ref{eq:Vecc}), and analogously to
Eq.~(\ref{eq:bisepl}) we may write
\begin{equation}
{\cal\breve B}^l_m =\sum_{\beta=0}^{m-1}\left|\sum_{\gamma=0}^{m-1}\tilde
M_{\beta\gamma} \left(\begin{array}{c}\cos\frac{\pi\gamma}{m}\\
\sin\frac{\pi\gamma}{m}\end{array}\right)\right|.
\label{eq:biseplns}
\end{equation}
For $\beta=0$, $\tilde M_{0\gamma}=1$, and we have to sum $m$ unit
vectors pointing towards consecutive corners of a polygon of $2m$
sides, the usual formation, the length of the resulting vector is
$1/\sin(\pi/2m)$. For $\beta=1$ only the last element of the row
will be $-1$. But if we change the sign of just the last vector of
the formation, we get the same formation rotated by an angle of
$\pi/2m$. This formation will give the same result. The next line
will give a formation rotated further by $\pi/2m$, and so on,
therefore the result is $m/\sin(\pi/2m)$. This is a lower bound
for the biseparable value, which is equal to the upper bound if
$m$ is a power of 2. In this case the ratio of the quantum and the
biseparable limits is ${\cal\breve Q}_m/{\cal\breve B}_m=m\sin(\pi/2m)$, resulting in the threshold visibility $V={\cal\breve B}_m/{\cal\breve Q}_m=1/(m\sin(\pi/2m))$. This is the same visibility obtained under Eq.~(\ref{eq:QperB}) by means of the Bell inequality~(\ref{eq:Bellcont}) of section~\ref{ketto}.

However, we would like to mention that this family of inequalities is more economical than our previous one. Namely, the number of joint measurements involved in Bell inequality~(\ref{eq:Bellcont}) scales as $m^3$, whereas the present Bell inequality defined by~(\ref{eq:BellPT}) consists of only $m^2$ joint measurements. Even for smaller number of measurements, the case which is more relevant to experiments, the difference is not negligible: Inequality~(\ref{eq:Bellcont}) (or equivalently the Bancal et al.~inequality~\cite{BGLP}) gives the threshold visibility $V=0.666$, requiring 18 joint correlation terms. On the other hand, the inequality defined by~(\ref{eq:BellPT}) yields the lower threshold $V=0.653$, using only 16 joint terms.

\section{Conclusion}

In this paper we extended the three-party three-setting inequality of Bancal et al.~\cite{BGLP}, which serves as a device-independent genuine tripartite entanglement witness, to an arbitrary number of settings. Our Bell inequalities (see Eq.~(\ref{eq:Bellcont}) and Eq.~(\ref{eq:BellPT}) for their definitions) can detect genuine tripartite entanglement in the noisy 3-qubit GHZ state with a visibility threshold of $V=1/(m\sin(\pi/2m))$, where $m$ denotes the number of settings per party. For $m=2,3$ our result recovers the threshold values corresponding to the Mermin inequality~\cite{Mermin} and the Bancal et al.~inequality~\cite{BGLP}, respectively. Numerical optimization suggests that these threshold values are optimal for $m=2$ and $m=3$. However, it is still an open question whether our family of inequalities (defined by Eq.~(\ref{eq:Bellcont})) is optimal for any value of $m$. The optimality of these inequalities for $m>3$ is supported by the fact that for $m$ going to infinity the visibility $V$ approaches the value of $2/\pi$, achievable by a biseparable model simulating three-party correlators. Also, it would be desirable to generalize our families either to more parties (here the method of Appendix~C in Ref.~\cite{BGLP} might be instructive) or to more outcomes. One may also wonder whether the inequalities presented in this work are optimal for important states different from the 3-qubit GHZ state. Furthermore, it would be of interest to find a Bell inequality, which is not the sum of three-party-correlators, giving a threshold visibility lower than $2/\pi$ for the noisy 3-qubit GHZ state.

\vspace{10pt}
\noindent\textbf{Acknowledgements} T.V. has been supported by a
J\'anos Bolyai Programme of the Hungarian Academy of Sciences.

\end{document}